\documentclass[aps,pra,amsmath,amssymb,reprint,groupedaddress,showpacs]{revtex4-1}

\usepackage{graphicx}% Include figure files
\usepackage{dcolumn}% Align table columns on decimal point
\usepackage{bm}% bold math

\begin{document}

% Use the \preprint command to place your local institutional report
% number in the upper righthand corner of the title page in preprint mode.
% Multiple \preprint commands are allowed.
% Use the 'preprintnumbers' class option to override journal defaults
% to display numbers if necessary
%\preprint{}

%Title of paper
\title{Ground state and the spin precession of the Dirac electron in counterpropagating plane electromagnetic waves}

% repeat the \author .. \affiliation  etc. as needed
% \email, \thanks, \homepage, \altaffiliation all apply to the current
% author. Explanatory text should go in the []'s, actual e-mail
% address or url should go in the {}'s for \email and \homepage.
% Please use the appropriate macro foreach each type of information

% \affiliation command applies to all authors since the last
% \affiliation command. The \affiliation command should follow the
% other information
% \affiliation can be followed by \email, \homepage, \thanks as well.
\author{G. N. Borzdov}
%\email[]{Your e-mail address}
\email[]{BorzdovG@bsu.by}
%\homepage[]{Your web page}
%\thanks{}
%\altaffiliation{}
\affiliation{Department of Theoretical Physics and Astrophysics, Belarusian State University,
4 Nezavisimosti Av., 220030 Minsk, Belarus}

%Collaboration name if desired (requires use of superscriptaddress
%option in \documentclass). \noaffiliation is required (may also be
%used with the \author command).
%\collaboration can be followed by \email, \homepage, \thanks as well.
%\collaboration{}
%\noaffiliation

%\date{\today}

\begin{abstract}
% insert abstract here
The fundamental solution of the Dirac equation for an electron in an electromagnetic field with harmonic dependence on space-time coordinates is obtained. The field is composed of three standing plane harmonic waves with mutually orthogonal phase planes and the same frequency. Each standing wave consists of two eigenwaves with different complex amplitudes and opposite directions of propagation. The fundamental solution is obtained in the form of the projection operator defining the subspace of solutions to the Dirac equation. It is illustrated by the analysis of the ground state and the spin precession of the Dirac electron in the field of two counterpropagating plane waves with left and right circular polarizations. Interrelations between the fundamental solution and approximate partial solutions is discussed and a criterion for evaluating accuracy of approximate solutions is suggested.
\end{abstract}

% insert suggested PACS numbers in braces on next line
\pacs{03.65.Pm, 03.30.+p, 02.30.Nw,  02.30.Tb}
% insert suggested keywords - APS authors don't need to do this
%\keywords{}

%\maketitle must follow title, authors, abstract, \pacs, and \keywords
\maketitle

% body of paper here - Use proper section commands
% References should be done using the \cite, \ref, and \label commands
\section{Introduction}
% Put \label in argument of \section for cross-referencing
%\section{\label{}}
Considerable recent attension has been focussed on the possibility of time and space-time crystals~\cite{qtime,cltime,ions,gaps}, analogous to ordinary crystals in space. The papers~\cite{qtime,cltime} provide the affirmative answer to the question, whether time-translation symmetry might be spontaneously broken in a closed quantum-mechanical system~\cite{qtime} and a time-independent, conservative classical system~\cite{cltime}. A space-time crystal of trapped ions and a method to realize it experimentally by confining ions in a ring-shaped trapping potential with a static magnetic field is proposed in~\cite{ions}. Standing electromagnetic waves comprize another type of space-time crystals. It was shown~\cite{gaps} that one can treat the space-time lattice, created by a standing plane electromagnetic wave, by analogy with the crystals of nonrelativistic solid state physics. In particular, the wave functions, calculated within this framework by using the first-order perturbation theory for the Schr\"{o}dinger-Stuekelberg equation, are Bloch waves with energy gaps~\cite{gaps}. The analylitical solution for the Kein-Gordon equation in the case of a field composed of two counterpropagating laser waves is obtained in~\cite{Hu2015}.

Standing electromagnetic waves constitute an interesting family of localized fields which may have important practical applications. In particular, optical standing waves can be used to focus atoms and ions onto a surface in a controlled manner, nondiffracting Bessel beams can be used as optical tweezers which are noninvasive tools generating forces powerful enough to manipulate microscopic particles. Superpositions of homogeneous plane waves propagating in opposite directions, the so-called Whittaker expansions, play a very important role in analyzing and designing localized solutions to various homogeneous partial differential equations \cite{Don92,*Don93,*Sha95}.

In this article we treat the motion of the Dirac electron in an electromagnetic field with four-dimensional periodicity, i.e., with periodic dependence on all four space-time coordinates. In terms of the three-dimensional description, such electromagnetic space-time crystal (ESTC) can be treated as a time-harmonic 3D standing wave. In solid state physics, the motion of electrons in natural crystals is described by the Schr\"{o}dinger equation with a periodic electrostatic scalar potential. The description of the motion of electrons in ESTCs by the Dirac equation takes into account both the space-time periodicity of the vector potential and the intrinsic electron properties (charge, spin, and magnetic moment). In  this case, the Dirac equation reduces to an infinite system of matrix equations. To solve it, we generalize the operator methods developed in~\cite{cr1,*cr2,*cr3,*oc92,*jmp93,*jmp97,*Wiley} to the cases of infinite-dimensional spaces and finite-dimensional spaces with any number of space dimensions. The evolution, projection and pseudoinverse operators are of major importance in this approach. The evolution operator (the fundamental solution of a wave equation) describes the field dependence on the space-time coordinates for the whole family of partial solutions. The method of projection operators is very useful at problem solving in classical and quantum field theory~\cite{Fed58,Fed,Bogush}. It was developed by Fedorov~\cite{Fed58,Fed} to treat finite systems of linear homogeneous equations. In the frame of Fedorov's approach, it is necessary first to find projection operators which define subspaces of solutions for two subsystems (constituent parts) of the system to solve, and then to find its fundamental solution, i.e., the projection operator defining the intersection of these subspaces, by calculating the minimal polynomial for some Hermitian matrix of finite dimensions. We present a different approach, based on the use of pseudoinverse operators, which is applicable to both finite and infinite systems of equations and has no need of minimal polynomials.

The fundamental solution of the Dirac equation for the field composed of three standing plane harmonic waves with mutually orthogonal phase planes and the same frequency is presented in Sec.~\ref{sec:baseq}. The case of two counterpropagating plane waves with left and right circular polarizations is treated in Sec.~\ref{sec:two}. Additional information on the numerical implementation of the presented approach and some results of its computer simulation can be found in~\cite{ESTCp1,ESTCp2,ESTCp3}.

\section{\label{sec:baseq}Basic relations}
\subsection{Matrix form}
An electron in an electromagnetic field with the four-dimensional potential $\bm{A}=(\textbf{A},i \varphi)$ is described by the Dirac equation
\begin{equation}\label{Dirac}
   \left[\gamma_k\left(\frac{\partial}{\partial x_k} - i A_k\frac{e}{c \hbar}\right) + \kappa_{e}\right]\Psi=0,
\end{equation}
where $\kappa_{e}=m_{e} c/ \hbar$, $c$ is the speed of light in vacuum, $\hbar$ is the Planck constant, $e$ is the electron charge, $m_{e}$ is the electron rest mass,  $\gamma_k$ are the Dirac matrices, $\Psi$ is the bispinor, $x_1$, $x_2$ and $x_3$  are the Cartesian coordinates, $x_4=i c t$, and summation over repeated indices is carried out from 1 to 4. In~\cite{ESTCp1,ESTCp2,ESTCp3} we have treated the field with $A_4\equiv i \varphi=0$ and
\begin{equation}\label{field}
  \textbf{A}^\prime \equiv \frac{e}{m_e c^2}\textbf{A}=\sum_{j=1}^6\left(\textbf{A}_j e^{i \bm{K}_j \cdot \bm{x}}+\textbf{A}_j^\ast e^{-i \bm{K}_j \cdot \bm{x}}\right),
\end{equation}
which is composed of six plane waves with unit wave normals $\pm\textbf{e}_\alpha$, where $\textbf{e}_\alpha$ are the orthonormal basis vectors, $\alpha =1, 2, 3$; $\bm{x}=(\textbf{r},ict)$, $\textbf{r}=x_1\textbf{e}_1+x_2\textbf{e}_2+x_3\textbf{e}_3$. All six waves
have the same frequency $\omega_0$ and
\begin{eqnarray}
\bm{K}_j&=&k_0 \bm{N}_j, \quad j=1,2,...,6, \quad k_0=\frac{\omega_0}{c}=\frac{2\pi}{\lambda_0},\nonumber
\\
\bm{N}_j&=&(\textbf{e}_j,i), \quad \bm{N}_{j+3}=(-\textbf{e}_j,i), \quad j=1,2,3. \label{KNj}
\end{eqnarray}
They may have any polarization, so that their complex amplitudes are specified by dimensionless real constants  $a_{jk}$ and $b_{jk}$ as follows
\begin{equation}
    \textbf{A}_j = \sum_{k=1}^3 \left(a_{jk} + i b_{jk}\right)\textbf{e}_k, \quad j = 1,2,...,6,\label{abjk}
\end{equation}
where $a_{jj} = b_{jj} = a_{j+3\,j} = b_{j+3\,j} = 0, j=1,2,3.$

For the electromagnetic lattice under consideration, the solution of Eq.~(\ref{Dirac}) can be found in the form of a Fourier series
\begin{equation}\label{sol1}
    \Psi=\Psi_0 e^{i \bm{x}\cdot\bm{K}},\quad \Psi_0=\sum_{n \in \mathcal L} c(n) e^{i \bm{x}\cdot\bm{G}(n)},
\end{equation}
where $\bm{K} = (\textbf{k},i \omega /c)$ is the four-dimensional wave vector, $\textbf{k} = k_1\textbf{e}_1+k_2\textbf{e}_2+k_3\textbf{e}_3$, $n=(n_1,n_2,n_3,n_4)$ is the multi-index specifying  $\textbf{n} = n_1\textbf{e}_1+n_2\textbf{e}_2+n_3\textbf{e}_3$ and $\bm{G}(n) = (k_0 \textbf{n},i k_0 n_4)$. Here, $c(n)$ are the Fourier amplitudes (bispinors), and $\mathcal L$ is the infinite set of all multi-indices $n$ with an even value of the sum $n_1+n_2+n_3+n_4$. Substitution of $\textbf{A}$ (\ref{field}) and $\Psi$ (\ref{sol1}) in Eq.~(\ref{Dirac}) results in the infinite system of matrix equations
\begin{equation}\label{meq}
\sum_{s \in S_{13}} V(n,s)c(n+s)=0, \quad {n\in \mathcal{L}},
\end{equation}
where
\begin{eqnarray}
% \nonumber to remove numbering (before each equation)
 S_{13}=&&\{s_h(i),i=0,1,...,12\}\label{S13}\\
 =&&\left\{(0, 0, 0, 0), \right. \nonumber\\
 &&(0, 0, -1, -1),(0, -1, 0, -1),(-1, 0, 0, -1),\nonumber\\
 &&(1, 0, 0, -1),(0, 1, 0, -1),(0, 0, 1, -1),\nonumber\\
 &&(0, 0, -1, 1),(0, -1, 0, 1),(-1, 0, 0, 1),\nonumber\\
 &&\left. (1, 0, 0, 1),(0, 1, 0, 1),(0, 0, 1, 1)\right\}\nonumber
\end{eqnarray}
is the set of 13 values of function $s_h=s_h(i)$, where $s_h(0)=(0,0,0,0)$ is the null shift. At $i=1,...,12$, this function specifies the shifts $s=(s_1,s_2,s_3,s_4)=s_h(i)$ of multi-indices $n$, defined by the Fourier spectrum of the field $\textbf{A}$ (\ref{field}), which satisfy the condition $|s_1|+|s_2|+|s_3|=|s_4|=1$. Because of this, they will be denoted the shifts of the first generation [$g_{4d}(s)=1$]. By the definition, $g_{4d}(s_1,s_2,s_3,s_4)=\max\{|s_1|+|s_2|+|s_3|,|s_4|\}$. Thus, each equation of the system relates 13 Fourier amplitudes (bispinors), in other words, each amplitude enters in 13 different matrix equations. We intensively use indexing of various mathematical objects by points $n=(n_1,n_2,n_3,n_4)$ of the integer lattice $\mathcal{L}$. The sequential numbering of these points, based on the use of $g_{4d}(n)$, drastically simplifies both numerical implementation of the presented techniques and analysis of solutions, because it takes into account the specific Fourier spectra of the electromagnetic lattice and the wave function, as well the structure of the finite models described below and in more detail in~\cite{ESTCp2}.

It is well known, e.g., see Ref.~\cite{Axi}, that 16 Dirac matrices form a basis in the space of $4\times 4$ matrices. In appendix, we present a specific numeration of these basis matrices $\Gamma_k, k=0,...,15$, which makes it possible, in particular, to reconstruct any matrix $\Gamma_k$ from its number $k$, see Ref.~\cite{ESTCp1}. Any $4\times 4$ matrix $V=\sum_{k=0}^{15}V_k\Gamma_k$ is uniquely defined by the set of its components $D_s(V)=\{V_k\}$ in the Dirac basis [Dirac set of matrix $V$, briefly, D-set of $V$]. Due to the structure of the Dirac equation, such expansions yield a convenient way to represent derived matrix expressions in a concise form, accelerate numerical calculations, and reduces data files. This approach is of particular assistance in solving the system of Eqs.~(\ref{meq}), see Ref.~\cite{ESTCp1,ESTCp2,ESTCp3}. D-sets of matrices $V(n,s)$ are presented in~\cite{ESTCp1} as functions of the dimensionless parameters
\begin{equation}\label{Qw}
    \bm Q = (\textbf{q},i q_4) = {\bm K}/\kappa_e, \quad  \Omega = \frac{\hbar \omega_0}{m_e c^2},
\end{equation}
\begin{equation}\label{qq4}
\textbf{q} = q_1\textbf{e}_1+q_2\textbf{e}_2+q_3\textbf{e}_3 = \frac{\hbar \textbf{ k}}{m_e c}, \quad   q_4 = \frac{\hbar \omega}{m_e c^2}.
\end{equation}

\subsection{Operator form}
Let us treat the infinite set $C =\{c(n),n \in {\mathcal L}\}$ of the Fourier amplitudes $c(n)$ of the wave function $\Psi$ (\ref{sol1}) as an element of an infinite dimensional linear space $V_C$. Since, for any $n \in {\mathcal L}$,
\begin{equation}\label{cn}
    c(n) = \left(
             \begin{array}{c}
               c^1(n) \\
               c^2(n) \\
               c^3(n) \\
               c^4(n) \\
             \end{array}
           \right) \equiv \left(
           \begin{array}{c}
               c^1 \\
               c^2 \\
               c^3 \\
               c^4 \\
             \end{array}
           \right)_n
\end{equation}
is the bispinor, $C \in V_C$ will be denoted the multispinor. Let us define a basis $e_j(n)$ in $V_C$ and the dual basis $\theta^j(n) = e_j^{\dag}(n)$ in the space of one-forms $V_C^\ast$ ($n \in {\mathcal L}$):
\begin{eqnarray}\label{e14n}
% \nonumber to remove numbering (before each equation)
  e_1(n)&=&\left(
             \begin{array}{c}
               1 \\
               0 \\
               0 \\
               0 \\
             \end{array}
           \right)_n, \quad e_2(n) = \left(
             \begin{array}{c}
               0 \\
               1 \\
               0 \\
               0 \\
             \end{array}
           \right)_n, \nonumber\\
  e_3(n)&=&\left(
             \begin{array}{c}
               0 \\
               0 \\
               1 \\
               0 \\
             \end{array}
           \right)_n, \quad e_4(n) = \left(
             \begin{array}{c}
               0 \\
               0 \\
               0 \\
               1 \\
             \end{array}
           \right)_n,
\end{eqnarray}
\begin{eqnarray}\label{t14n}
% \nonumber to remove numbering (before each equation)
  \theta^1(n)&=&\left(
                          \begin{array}{cccc}
                            1 & 0 & 0 & 0 \\
                          \end{array}
                        \right)_n, \quad \theta^2(n) = \left(
                          \begin{array}{cccc}
                            0 & 1 & 0 & 0 \\
                          \end{array}
                        \right)_n,\nonumber  \\
  \theta^3(n)&=&\left(
                          \begin{array}{cccc}
                            0 & 0 & 1 & 0 \\
                          \end{array}
                        \right)_n, \quad \theta^4(n) = \left(
                          \begin{array}{cccc}
                            0 & 0 & 0 & 1 \\
                          \end{array}
                        \right)_n .
\end{eqnarray}
In this notation, the system of equations (\ref{meq}) takes the form
\begin{equation}\label{fC}
    \langle f^j(n),C\rangle \equiv \sum_{s \in S_{13}} V^j{}_k(n,s) c^k(n+s) = 0,
\end{equation}
where $j=1,2,3,4$, $n \in {\mathcal L}$, and
\begin{eqnarray}
% \nonumber to remove numbering (before each equation)
  &&f^j(n)=\sum_{s \in S_{13}} V^j{}_k(n,s) \theta^k(n+s),\nonumber\\
  &&\langle f^j(n),e_k(n+s)\rangle=V^j{}_k(n,s).\label{fjn}
\end{eqnarray}
These relations can be rearranged to the basic system of equations
\begin{equation}\label{PnC}
    P(n)C = 0, \quad n \in {\mathcal L},
\end{equation}
where
\begin{equation}\label{Pnfaf}
    P(n) = [f^{\alpha}(n)]^\dag \otimes a^{\alpha}{}_\beta (n)f^\beta (n)
\end{equation}
is the Hermitian projection operator with the trace $tr[P(n)]=4$ and following properties:
\begin{equation}\label{Pn2}
   [P(n)]^2 = [P(n)]^\dag= P(n),
\end{equation}
\begin{equation}\label{anL}
    a(n) = [L(n)]^{-1}, \quad L^\alpha{}_\beta (n) =\left\langle f^{\alpha}(n),\left[f^{\beta}(n)\right]^\dag\right\rangle,
\end{equation}
where $\alpha , \beta=1,2,3,4$. The Hermitian $4\times 4$ matrices $L(n)$ and $a(n)$ at $n=(n_1,n_2,n_3,n_4)$ are defined by the following D-sets:
\begin{eqnarray}
   &&D_s[L(n)]=\left\{1+I_A+w_1^2+w_2^2+w_3^2+w_4^2,0,0,0,\right.\nonumber\\
   &&\left. -2w_4,0,0,0,0,2w_3w_4,2w_1w_4,2w_2w_4,0,0,0,0 \right\},\label{dsLn}
\end{eqnarray}
\begin{eqnarray}
   &&D_s[a(n)]=\frac{1}{\left|L(n)\right|}\left\{1+I_A+w_1^2+w_2^2+w_3^2+w_4^2,0,0,0,\right.\nonumber\\
   &&\left. 2w_4,0,0,0,0,-2w_3w_4,-2w_1w_4,-2w_2w_4,0,0,0,0\right\},\label{dsan}
\end{eqnarray}
where
\begin{eqnarray}
    I_A =&& 2\sum_{j=1}^6 \left|\bm{A}_j \right|^2 = 2\left(a_{12}^2+b_{12}^2+a_{13}^2+b_{13}^2+a_{21}^2+b_{21}^2\right. \nonumber\\ &&+a_{23}^2+b_{23}^2+a_{31}^2+b_{31}^2+a_{32}^2+b_{32}^2\nonumber \\
    &&+a_{42}^2+b_{42}^2+a_{43}^2+b_{43}^2+a_{51}^2+b_{51}^2\nonumber \\
    &&\left. +a_{53}^2 +b_{53}^2+a_{61}^2+b_{61}^2+a_{62}^2+b_{62}^2\right),\label{IA}
\end{eqnarray}
\begin{eqnarray}
% \nonumber to remove numbering (before each equation)
  \left|L(n)\right|=&&I_A^2 + 2I_A\left(1+w_1^2+w_2^2+w_3^2+w_4^2\right)\nonumber\\
  &&+\left(1+w_1^2+w_2^2+w_3^2-w_4^2\right)^2,\label{dLn}
\end{eqnarray}
and  $w_j=q_j+n_j \Omega$. It is significant that, for a nonvanishing electromagnetic field ($I_A\neq 0$), the determinant $\left|L(n)\right|>0$ and hence equations (\ref{PnC})--(\ref{dLn}) are valid for any $n \in {\mathcal L}$.

\subsection{Fundamental solution}
The fundamental solution  $\mathcal{S}$, i.e., the operator of projection onto the solution subspace of the multispinor space $V_C$, and the projection operator $\mathcal{P}$ of the infinite system of equations~(\ref{PnC}) are defined as follows~\cite{ESTCp1}
\begin{equation}\label{Pro}
  \mathcal{S}=\mathcal{U}-\mathcal{P},\quad  \mathcal{P}=\sum_{k=0}^{+\infty}\sum_{n\in\mathcal{F}_k}\rho_k(n),
\end{equation}
\begin{equation}\label{FkL}
    \bigcup_{k=0}^{+\infty}\mathcal{F}_k=\mathcal{L}, \quad \mathcal{F}_j\bigcap\mathcal{F}_k=\emptyset, \quad j\neq k,
\end{equation}
where $\rho_k(n)$ are Hermitian projection operators with the trace $tr[\rho_k(n)]=4$, and $\mathcal{U}$ is the unit operator in $V_C$, which can be written as
\begin{equation}\label{UVC}
    \mathcal{U}=\sum_{n\in \mathcal{L}} I(n),\quad I(n)=e_j(n)\otimes\theta^j(n), \quad tr[I(n)]=4.
\end{equation}
For any $C_0\in V_C$, $C=\mathcal{S}C_0$ is a partial solution of Eq.~(\ref{PnC}), i.e., the function $\Psi$~(\ref{sol1}) with the set of Fourier amplitudes $\{c(n), n \in \mathcal{L}\}=\mathcal{S}C_0$  satisfies the Dirac equation~(\ref{Dirac}) for the problem under consideration.

The Hermitian operator $\mathcal P$ of the system of equations~(\ref{PnC}), by definition (see appendix), has the following properties
\begin{equation}\label{Pdag}
    \mathcal{P}^\dag=\mathcal{P}^2=\mathcal{P}, \quad P(n)\mathcal{P}=\mathcal{P}P(n)=P(n)
\end{equation}
for any $n\in{\mathcal L}$, and $\rho_k(n)$ satisfy the relations
\begin{equation}\label{rok}
    \rho_k^{\dag}(n)=\rho_k^2(n)=\rho_k(n), \quad tr[\rho_k(n)]=4, \quad n \in \mathcal{L},
\end{equation}
\begin{equation}\label{romn}
    \rho_k(m)\rho_l(n)=0 \text{ if } k\neq l \text{ or (and) } m\neq n,
\end{equation}
\begin{equation}\label{ro0}
    \rho_0(n)=P(n), \quad n\in \mathcal{F}_0.
\end{equation}
There exist various ways to split the lattice $\mathcal L$ into sublattices $\mathcal{F}_k$ to fulfil conditions (\ref{FkL}) and (\ref{romn}), one of them is described in Ref.~\cite{ESTCp2}. Providing these conditions are met, substitution of
\begin{equation}\label{albero}
    \alpha=\sum_{j=0}^{k-1}\sum_{n\in \mathcal{F}_j}\rho_j(n)\equiv\mathcal{P}_{k-1}, \quad \beta=P(m), \quad m\in \mathcal{P}_{k}
\end{equation}
into Eqs.~(\ref{Aad}) and (\ref{gam}) results in $\rho_k(m)=\delta$ (\ref{Aad}).

It follows from Eq.~(\ref{Pnfaf}) that
\begin{equation}\label{PmPn}
    P(m)P(n)=\left[f^i(m)\right]^{\dag}\otimes\left[a(m)N(m,n)a(n)\right]^i{}_j f^j(n),
\end{equation}
where
\begin{equation}\label{Nij}
    N^i{}_j(m,n)=\left\langle f^i(m),\left[f^j(n)\right]^{\dag}\right\rangle,\quad i,j=1,2,3,4,
\end{equation}
$N(n,n)\equiv L(n)$ (\ref{anL}). At any given n, Eq.~(\ref{meq}) relates the Fourier amplitude $c(n)$ only with 12 amplitudes $c(n+s)$, where $g_{4d}(s)=1$. In consequence of this, $N(m,n)\equiv 0$ at $g_{4d}(n-m)>2$. Substitution of (\ref{fjn}) in (\ref{Nij}) at $n=m+s$ gives
\begin{eqnarray}\label{N1N2}
% \nonumber to remove numbering (before each equation)
  N^{\dag}(n,m)=N(m,n)&=&L(m) \text{ for } n=m,\nonumber\\
                      &=&N_1(m,s) \text{ for } g_{4d}(s)=1,\nonumber\\
                      &=&N_2(s)\Gamma_0 \text{ for } g_{4d}(s)=2.
\end{eqnarray}
The D-sets of 12 matrices $N_1(m,s)$ and the table of 56 scalar coefficients $N_2(s)$ are presented in Ref.~\cite{ESTCp3}. These major structural parameters of the electromagnetic lattice specify interrelations in the system of equations~(\ref{meq}).

The relations presented in this section and appendix provide convenient means to find operators $\rho_k(n)$ by making use the recurrent algorithm devised to minimize volumes of computations and data files~\cite{ESTCp1,ESTCp2}. It begins with the selection of an infinite subsystem consisting from independent equations and the calculation of the projection operators $\rho_0(n)=P(n), \quad n\in \mathcal{F}_0 \subset \mathcal{L}$, which uniquely define the fundamental solutions of these equations. At each new step of the recurrent process, we add another infinite set of mutually independent equations which, however, are related with some of the equations introduced at the previous steps. Consequently, we obtain an infinite set of independent finite systems of interrelated equations [fractal clusters of equations]. It can be described as a 4d lattice of such clusters. Each step of the recurrent procedure expands clusters for which it provides the exact fundamental solutions.

\subsection{\label{sec:approx}Approximate solutions}
Numerical implementation of the obtained solution implies the replacement of the projection operator $\mathcal{P}$ (\ref{Pro}) of the infinite system~(\ref{PnC}) by the projection operator
\begin{equation}\label{Pprime}
    \mathcal{P'}=\sum_{k\in k_L}\sum_{n\in n_L(k)}\rho_k(n)
\end{equation}
of its finite subsystem
\begin{equation}\label{PkbLC}
    P(n)C=0,\quad n\in \mathcal{L}'=\bigcup_{k\in k_L}n_L(k) \subset \mathcal{L}.
\end{equation}
Here, $k_L$ is an ordered finite list of integers, and $n_L(k)$ is a finite list of points $n\in\mathcal{F}_k$, specifying a finite model of the infinite lattice. The projection operator
\begin{equation}\label{SkbL}
    \mathcal{S'} = \mathcal{U} - \mathcal{P'}
\end{equation}
defines the exact fundamental solution of Eq.~(\ref{PkbLC}), which is also the approximate solution of Eq.~(\ref{PnC}), provided by this finite model.

In this article, we restrict our consideration to the case when the amplitude $C_0$ specifying a partial solution is given by $C_0=a_0^j e_j(n_o)$, $n_o=(0,0,0,0)$, and the relation
\begin{equation}\label{CSd}
    C=\{c(n),n\in S_d\}=\mathcal{S'}C_0=C_0-\mathcal{P'}C_0
\end{equation}
describes the four-dimensional subspace of exact solutions of Eq.~(\ref{PkbLC}). Here, $S_d\subset\mathcal{L}$ is the solution domain, i.e., the subset of $\mathcal{L}$ with nonzero bispinors $c(n)$. Bispinors $c(n)$ and $a_0$ are linearly related as
\begin{equation}\label{Snc0}
    c(n)=S(n)a_0,
\end{equation}
where $S(n)$ is the $4\times 4$ matrix, $S^i{}_j(n)=\left\langle\theta^i(n),\mathcal{S'}e_j(n_o)\right\rangle$ are defined in~\cite{ESTCp2}.
Substituting $c(n)$ in Eq.~(\ref{sol1}) gives
\begin{equation}\label{Exc0}
    \Psi=\sum_{n\in S_d}c(n)e^{i \varphi_n(\bm{x})}\equiv E_v a_0,
\end{equation}
where
\begin{equation}\label{Ex}
    E_v=\sum_{n\in S_d}e^{i \varphi_n(\bm{x})}S(n)
\end{equation}
is the evolution operator. In terms of the dimensionless coordinates $\textbf{r}'=\textbf{r}/\lambda_0=X_1\textbf{e}_1+X_2\textbf{e}_2+X_3\textbf{e}_3$, $X_4=ct/\lambda_0$,
the phase function $\varphi_n(\bm{x})$ can be written as
\begin{eqnarray}
% \nonumber to remove numbering (before each equation)
  \varphi_n(\bm{x})&=&(\textbf{k}+k_0\textbf{n})\cdot\textbf{r} - (\omega+\omega_0n_4)t \nonumber\\
   &=&2\pi\left[(\textbf{n}+\textbf{q}/\Omega)\cdot\textbf{r}' - (n_4+q_4/\Omega)X_4\right].\label{phinx}
\end{eqnarray}

The evolution operator $E_v$ is the major characteristic of the whole family of partial solutions $\Psi$~(\ref{Exc0}). In particular, it provides a convenient way to calculate the mean value
\begin{equation}\label{meanA}
    \langle{A}\rangle=\frac{a_0^{\dag}A_E a_0}{a_0^{\dag}U_Ea_0}
\end{equation}
of an operator $A$ with respect to function  $\Psi$, where $A_E={\mathcal I}_{\Delta X}(E_v^{\dag}A E_v)$,
\begin{equation}\label{UE}
    U_E={\mathcal I}_{\Delta X}(E_v^{\dag} E_v)=\sum_{n\in Sd}S^{\dag}(n)S(n),
\end{equation}
\begin{equation}\label{IntX}
    {\mathcal I}_{\Delta X}(f)\equiv\int_{\Delta X}f dX_1dX_2dX_3dX_4,
\end{equation}
 and$\Delta X$ is the domain given by intervals $[X_k,X_k+1], k=1, 2, 3, 4.$

\subsection{\label{sec:accur}Evaluating accuracy of solutions}
The distinguishing feature of the presented technique is that each step of the recurrent procedure expands the subsystem of equations for which it provides the exact fundamental solution. One can check the calculation for accuracy by using relations~(\ref{rok}) and (\ref{romn}).  Substitution of $c(n)$~(\ref{Snc0}) into the left side of Eq.~(\ref{meq}) reduces it to the form $\mathcal{V}_S(n)a_0$, where
\begin{equation}\label{VSn}
    \mathcal{V}_S(n)=\sum_{s\in S_{13}}V(n,s)S(n+s).
\end{equation}
At $n\in \mathcal{L}'$, the equation $\mathcal{V}_S(n)a_0=0$ is satisfied at any $a_0$, because in this domain $\mathcal{V}_S(n)\equiv 0$. This provides means for final numerical checking of the fundamental solution $\mathcal{S'}$ of the system (\ref{PkbLC}) and the evolution operator $E_v(\bm{x})$~(\ref{Ex}) for accuracy~\cite{ESTCp2}.

Let $\mathcal{D}$ be a differential operator in a space $\mathcal{V}_{\Psi}$ of scalar, vector, spinor, or bispinor functions, and $\|\Psi\|$ be the norm of $\Psi$ on $\mathcal{V}_{\Psi}$. The functional
\begin{equation}\label{Rpsi}
    \mathcal{R}: \Psi\mapsto \mathcal{R}[\Psi]=\frac{\|\Psi_D\|}{\|\Psi\|}
\end{equation}
where $\Psi_D=\mathcal{D}\Psi$, evaluates the relative residual at the substitution of $\Psi$ into the differential equation $\mathcal{D}\Psi=0$. It provides a fitness criterion to compare in accuracy various approximate solutions of this equation. For an exact solution $\Psi$, the residual $\Psi_D$ vanishes, i.e., $\mathcal{R}[\Psi]=0$. If $\Psi_D\neq 0$, but $\mathcal{R}[\Psi]\ll 1$, the function $\Psi$ may be treated as a reasonable approximation to the exact solution, and the smaller is $\mathcal{R}[\Psi]$, the more accurate is the approximation. In terms of distances $d=\|\Psi\|$ and $d_D=\|\Psi_D\|$ of $\Psi$ and $\Psi_D$ to the origin of $\mathcal{V}_{\Psi}$ (the zero function), one can graphically describe $\mathcal{R}[\Psi]$  as shrinkage in distance $\mathcal{R}[\Psi]=d_D/d$. The functional $\mathcal{R}$, as applied to a family of functions $\Psi(\bm{x},\lambda)$ with members specified by a parameter $\lambda$, results in function $\mathcal{R}[\Psi(\bm{x},\lambda)]$ of $\lambda$, denoted below $\mathcal{R}(\lambda)$ for short.

To introduce this criterion in the problem under consideration, we first transform Eq.~(\ref{Dirac}) to the equivalent equation $\mathcal{D}\Psi=0$ with the dimensionless operator
\begin{equation}\label{Ddim}
    \mathcal{D}=\sum_{k=1}^3 \alpha_k\left(-\frac{i \hbar}{m_e c}\frac{\partial}{\partial x_k} - A'_k\right) - \frac{i \hbar}{m_e c^2}\frac{\partial}{\partial t} + \alpha_4.
\end{equation}
From Eqs.~(\ref{Exc0}) and (\ref{Ddim}) follows
\begin{equation}\label{psid}
    \Psi_{D}=\mathcal{D}\Psi=D_va_0,
\end{equation}
where $D_v=\mathcal{D}E_v$ is the evolution operator describing the family of remainder functions $\Psi_D$~\cite{ESTCp2}.
The norm of $\Psi_D$ (\ref{psid}) can be written as
\begin{equation}\label{npsid}
    \left\|\Psi_D\right\|=\sqrt{a_0^{\dag} U_Da_0},
\end{equation}
where the matrix $U_D$ is presented in~\cite{ESTCp2}. Thus, for the function $\Psi$~(\ref{Exc0}), from the definition~(\ref{Rpsi}) follows
\begin{equation}\label{Rc0}
   \mathcal{R}=\sqrt{\frac{a_0^{\dag} U_Da_0}{a_0^{\dag} U_Ea_0}}.
\end{equation}

\subsection{Orthogonality relation}
Let $\Psi_a=\Psi_{0a} e^{i \bm{x}\cdot\bm{K}_a}$ and $\Psi_b=\Psi_{0b} e^{i \bm{x}\cdot\bm{K}_b}$ be solutions of the Dirac equation, i.e., $\mathcal{D}\Psi_a\equiv 0, \mathcal{D}\Psi_b\equiv 0$, where $\bm{K}_a = (\textbf{k},i \omega_a /c), \bm{K}_b = (\textbf{k},i \omega_b /c), \omega_a\neq\omega_b$, and
\begin{equation}\label{psiab}
    \Psi_{0a}=\sum_{n \in \mathcal L} a(n) e^{i \bm{x}\cdot\bm{G}(n)},\quad \Psi_{0b}=\sum_{n \in \mathcal L} b(n) e^{i \bm{x}\cdot\bm{G}(n)}.
\end{equation}
Upon integrating the identity $\Psi_b^{\dag}\mathcal{D}\Psi_a-(\Psi_a^{\dag}\mathcal{D}\Psi_b)^*\equiv 0$ we obtain the orthogonality relation ${\mathcal I}_{\Delta X}(\Psi_{0b}^{\dag}\Psi_{0a})=0$, which can be also written as
\begin{equation}\label{orth0}
    \sum_{n \in \mathcal L} b^{\dag}(n)a(n)=0.
\end{equation}

\subsection{Dispersion relation}
It should be emphasized that the analytical fundamental solution $\mathcal{S}$~(\ref{Pro}) is obtained without recourse to any dispersion relation, i.e., for any vector $\bm Q$~(\ref{Qw}). Let us explain this on the example of the exact Volkov solution for an electron in the field of a plane wave. There exist different representations of this solution~\cite{Fed,Tern}. We present below another one which is more straightforward and convenient for our purposes. In this particular case, there is only one wave of six waves in Eq.~(\ref{field}), namely, the wave with amplitude $\textbf{A}_3 = a_{31}\textbf{e}_1 + i b_{32}\textbf{e}_2$.   Substituting $\Psi(\bm{x})=\Psi(\zeta)e^{i \kappa_{e} {\bm Q}\cdot \bm{x}}$ with $\zeta=\bm{N}_3\cdot\bm{x}=x_3-ct$ in Eq.~(\ref{Dirac}) gives an ordinary differential equation which has the exact solution $\Psi(\bm{x})=E_v(\bm{x})a_0$, where
\begin{equation}\label{exJ}
    E_v(\bm{x})=e^{i \Phi(\bm{x})}J(\zeta)
\end{equation}
is the evolution operator (the fundamental solution of this equation), $J=J(\zeta)$ is the $4\times 4$  projection matrix ($J^2=J, tr J=2$) defined by
\begin{eqnarray}
   &&D_s(J)=\left\{1/2,0,-i J_{11},i J_{10},J_4,0,0,0,\right.\nonumber\\
   &&\left. 0,-1/2,J_{10},J_{11},0,-i J_4,0,0 \right\},\label{dsJ}
\end{eqnarray}
\begin{eqnarray}
% \nonumber to remove numbering (before each equation)
  J_4 &=& [2(q_4-q_3)]^{-1},\, J_{10}=J_4(q_1-2 a_{31}\cos k_0\zeta),\nonumber \\
  J_{11} &=& J_4(q_2+2b_{32}\sin k_0\zeta).
\end{eqnarray}
At any given $\zeta$, the bispinor $\Psi(\zeta)$ belongs to the two-dimensional subspace defined by $J(\zeta)$. The phase function $\Phi$ consists of two parts which are linear in $\bm{x}$ and periodic in $\zeta$, respectively, as follows
\begin{eqnarray}
% \nonumber to remove numbering (before each equation)
  \Phi &=& \kappa_{e} {\bm Q}'\cdot\bm{x}+\frac{J_4}{\Omega}\left[4 b_{32}q_2(1-\cos k_0\zeta)\right.\nonumber\\
   &-& \left. 4a_{31}q_1\sin k_0\zeta +(a_{31}^2-b_{32}^2)\sin 2 k_0\zeta\right], \\
  {\bm Q}' &=& {\bm Q}-\frac{1+{\bm Q}^2+I_A}{2{\bm Q}\cdot\bm{N}_3}\bm{N}_3,\, I_A=2(a_{31}^2+b_{32}^2).
  \end{eqnarray}
It is easy to verify that ${\bm Q}'$ satisfies the dispersion relation $1+{\bm Q}'^2+I_A=0$ at any ${\bm Q}$. In other words, the fundamental solution has the build-in dispersion relation. Similarly, in optics of plane-stratified complex mediums, fundamental solutions (exponential evolution operators) define both wave vectors and polarizations of eigenwaves in an anisotropic or bianisotropic slab~\cite{cr1,*cr2,*cr3,*oc92,*jmp93,*jmp97,*Wiley}. It is convenient to preset ${\bm Q}$ satisfying the dispersion relation, then ${\bm Q}'\equiv{\bm Q}$ and the parameter $\xi_V=q_4-\sqrt{1+\textbf{q}^2}$ specifies the deviation from the free-space dispersion relation $1+\textbf{q}^2=q_4^2$ as follows
\begin{equation}\label{xiV}
    \xi_V=\sqrt{1+\textbf{q}^2+I_A}-\sqrt{1+\textbf{q}^2}
\end{equation}
for any given $\textbf{q}$.

In the general problem under study, the dispersion relation manifests itself in the spectral distribution of Fourier components $c(n)$~(\ref{sol1}). In numerical calculations for a finite model with a localized Fourier spectrum, when $ g_{4d}(n)\leq g_{max}$ for all $n$ in Eq.~(\ref{Pprime}), it has a pictorial presentation in the form of spectral curves of approximate solutions $\mathcal{R}_j=\mathcal{R}_j(\xi)$, where
\begin{equation}\label{xiq4}
    \xi=q_4-\sqrt{1+\textbf{q}^2}=\frac{\hbar \omega}{m_e c^2}-\sqrt{1+\left(\frac{\hbar \textbf{k}}{m_e c}\right)^2},
\end{equation}
and $\mathcal{R}_j=\sqrt{\lambda_j}$ is given by Eq.~(\ref{Rc0}) at $a_0=c_j$. The generalized eigenvalues $\lambda_j$ and eigenvectors $c_j$ are defined by the equation $U_Dc_j=\lambda_jU_Ec_j$ with the Hermitian $4\times 4$ matrices $U_E$ and $U_D$, and the quartic equation $\det(U_D -\lambda U_E)=0$ has real coefficients and positive roots $\lambda_j$ indexed below in increasing order of magnitude. The minimum  $\{\xi_0,\mathcal{R}_0=\mathcal{R}_1(\xi_0)\}$ of the spectral curve $\mathcal{R}_1=\mathcal{R}_1(\xi)$ specifies the most accurate approximate solution. It follows from the results of computer simulations~\cite{ESTCp3} that $\xi_0$ converges to a positive limit and $\mathcal{R}(\xi_0)$ tends to zero with increasing $g_{max}$. In the limit, $\Psi$~(\ref{Exc0}) converges to a family of exact solutions with the dispersion relation
\begin{equation}\label{dispeq}
    \frac{\hbar\omega}{m_e c^2}=\xi_0 + \sqrt{1+\left(\frac{\hbar\textbf{k}}{m_e c}\right)^2}.
\end{equation}

\section{\label{sec:two}Two counterpropagating waves}
\subsection{Dispersion relation}
\begin{figure}
\includegraphics{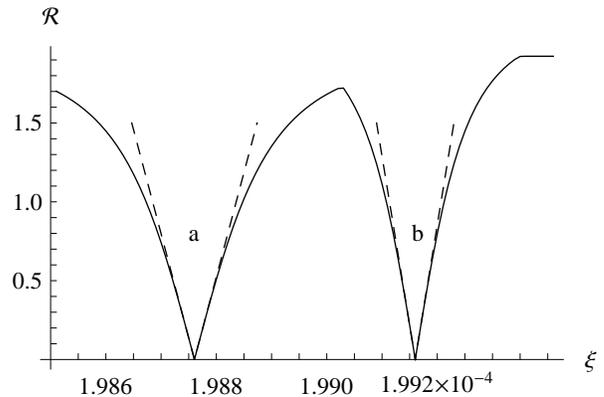}
\caption{\label{fig1}Spectral curve of approximate solutions $\mathcal{R}=\mathcal{R}_{1}(\xi)$ and its models $\mathcal{R}=\mathcal{R}_1^{ap}(\xi)$ (dashed curves) for the spectral lines (a) $\xi_0=\xi_{oa}=0.00019876$, $\mathcal{R}_{0}=1.77297\times 10^{-9}$, $\beta_0=1.32212\times 10^7$, $\delta\xi(\mathcal{R}_{av})=1.51272\times 10^{-9}$ and (b) $\xi_0=\xi_{ob}=0.00019916$, $\mathcal{R}_{0}=1.0835\times 10^{-9}$, $\beta_0=2.14323\times 10^7$, $\delta\xi(\mathcal{R}_{av})=9.33172\times 10^{-10}$ at $\Omega=0.1$, $\mathcal{R}_{av}=\sqrt{I_A}=0.02$, and $g_{max}=4$.}
\end{figure}

In this section we apply the presented technique to find the ground state of the Dirac electron with, by definition, the quasi-momentum $\textbf{p}=\hbar\textbf{k}=m_{e}c\textbf{q}=0$,  in the field of two counterpropagating circularly polarized waves with the same amplitude
\begin{equation}\label{A1A4}
    \textbf{A}_1=\textbf{A}_4=A_m(\textbf{e}_2+i \textbf{e}_3)/\sqrt{2}.
\end{equation}
The other four amplitudes in Eq.~(\ref{field}) are equal to zero and hence $I_A=4 A_m^2$. In this case, most of the structural parameters in Eq.~(\ref{N1N2})  are vanishing, only $N_1(m,s)$ for $s\in \{(-1,0,0,-1),(-1,0,0,1),(1,0,0,-1),(1,0,0,1)\}$ and $N_2(s)$ for $s\in \{(-2,0,0,0),(2,0,0,0)\}$ are not zero, therefore $\Psi$~(\ref{Exc0}) contains only Fourier components with $n=(n_1,0,0,n_4)$, where $|n_1|\leq 1+g_{max}$, whereas $|n_4|=0, 1$. Figure~\ref{fig1} shows the corresponding spectral curve of approximate solutions, which reveals that the ground state has two different frequency levels specified by minimums of spectral lines $a$ and $b$. Their bottom parts (see dash curves in Fig.~\ref{fig1}) can be closely approximated as follows
\begin{equation}\label{parabola}
    \mathcal{R}_1^{ap}(\xi)=\sqrt{\mathcal{R}_0^2+\beta_0^2(\xi-\xi_0)^2}.
\end{equation}
The half-width $\delta\xi(\mathcal{R}_{av})$ of the solution line, i.e., the half-width of $\xi$ domain, where $\mathcal{R}_0\leq\mathcal{R}\leq\mathcal{R}_{av}$, can be estimated from Eq.~(\ref{parabola}) as
\begin{equation}\label{dxi}
    \delta\xi(\mathcal{R}_{av})=\frac{1}{\beta_0}\sqrt{\mathcal{R}_{av}^2-\mathcal{R}_0^2}.
\end{equation}
This half-width is a rapidly decreasing function of $g_{max}$.

%%%%%%%%%%%%%%%%%%%%%%%%%%%%%figure1%%%%%%%%%%%%%%%
%%%%%%%%%%%%%%%%%%%%%%%%%%%%%%%%%%%%%%%%%%%%%%%%%%

The condition $\mathcal{R}_1\ll 1$ is satisfied within narrow limits of $\xi$ values, whereas $\mathcal{R}_{2,3,4}\gg\mathcal{R}_1$ and they do not satisfy the similar condition at any value of $\xi$, for example, $\{\mathcal{R}_j, j=2,3,4\}=\{1.92, 1.96, 2.12\}$ and $\{1.86, 1.92, 1.96\}$ at $\xi=\xi_{0a}$ and $\xi=\xi_{0b}$, respectively. Thus the amplitude subspaces in Eq.~(\ref{Exc0}) for both of levels are one-dimensional, they are specified by the generalized eigenvectors $a_{0a}=c_1(\xi_{0a})=a_{+}$ and $a_{0b}=c_1(\xi_{0b})=a_{-}$ or, in other words, by the projection matrices  $P_a=P_{+}$ and $P_b=P_{-}$, where
\begin{equation}\label{a0aa0b}
    a_{\pm}=\frac{1}{\sqrt{2}}\left(
             \begin{array}{c}
               \pm1 \\
               1 \\
               0 \\
               0 \\
             \end{array}
           \right), \quad P_{\pm}=\frac{1}{2}
             \left(
               \begin{array}{cccc}
                 1 & \pm 1 & 0 & 0 \\
                 \pm 1 & 1 & 0 & 0 \\
                 0 & 0 & 0 & 0 \\
                 0 & 0 & 0 & 0 \\
               \end{array}
             \right).
           \end{equation}

\begin{figure}
\includegraphics{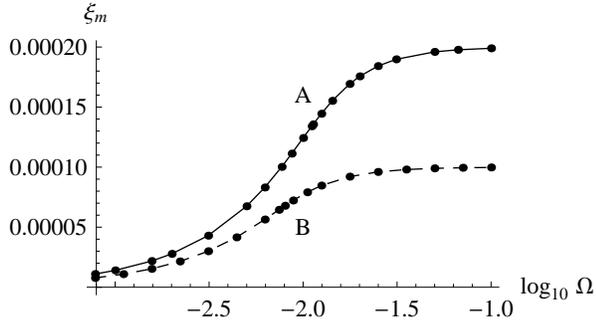}
\caption{\label{fig2}Plot of $\xi_m$ against $\log_{10}\Omega$ at (A) $I_A=0.0004$, and (B) $I_A=0.0002$.}
\end{figure}

\begin{figure}
\includegraphics{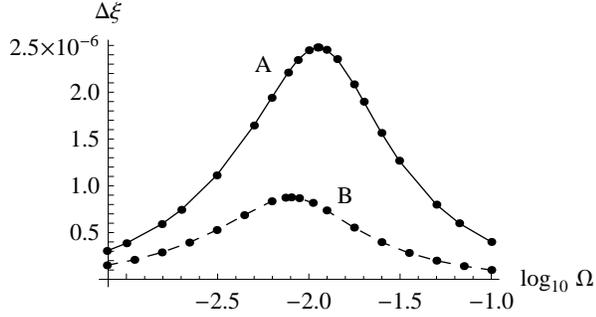}
\caption{\label{fig3}Plot of $\Delta\xi$ against $\log_{10}\Omega$ at (A) $I_A=0.0004$, and (B) $I_A=0.0002$.}
\end{figure}

\begin{figure}
\includegraphics{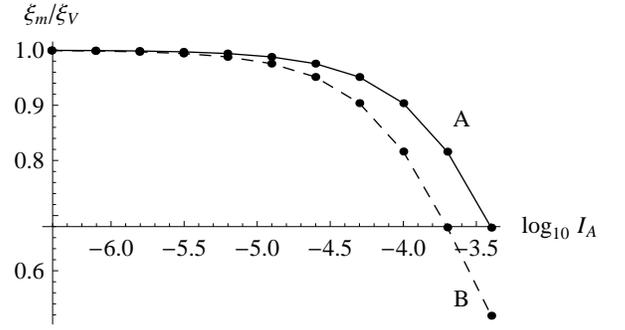}
\caption{\label{fig4}Ratio $\xi_m/\xi_V$ against $\log_{10} I_A$ at (A) $\Omega=\Omega_A$, and (B) $\Omega=\Omega_B$.}
\end{figure}

\begin{figure}
\includegraphics{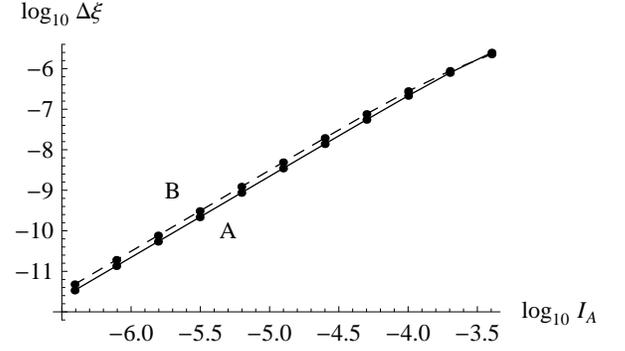}
\caption{\label{fig5}Plot of $\log_{10}\Delta\xi$ against $\log_{10} I_A$ at (A) $\Omega=\Omega_A$, and (B) $\Omega=\Omega_B$.}
\end{figure}

It is convenient to describe the closely spaced levels of the Dirac electron, i.e., the normalized frequencies $\xi_{0a}$ and $\xi_{0b}$, in terms of the mean value $\xi_m=\frac{1}{2}(\xi_{0a}+\xi_{0b})$ and the difference of levels $\Delta\xi=\xi_{0b}-\xi_{0a}$. The dependence of $\xi_m$ and $\Delta\xi$ on the normalized frequency $\Omega$ of the electromagnetic lattice is shown in Fiq.~\ref{fig2} and Fiq.~\ref{fig3}, respectively. The dots represent calculations, while the curves are obtained by the linear interpolation, for the range of $\Omega$ from 1/1280 to 1/10, i.e., for the X-ray standing waves with the wavelength $\lambda_0$ from 0.024 nm to 3.1 nm. In the central band of this range $\Delta\xi$ has the maximum at $\lambda_0=\lambda_A=0.2146318$ nm ($\Omega=\Omega_A=0.01130452$) and $\lambda_0=\lambda_B=0.3032564$ nm ($\Omega=\Omega_B=0.008000855$) for curves A and B in Fig.~\ref{fig3}, respectively. In the hard X-ray region $\xi_m$ weakly depends on $\Omega$, see Fiq.~\ref{fig2}. The dependence $\xi_m$ on $I_A$ can be approximated by $\xi_V=\sqrt{1+I_A}-1$ in a wide range of $I_A$, see Fiq.~\ref{fig4}. Figure~\ref{fig5} illustrates the dependence of $\Delta\xi$ on $I_A$ in this range. The smaller is $\Omega$ or the greater is $I_A$ or both, the greater is $g_{max}$ which provides reasonably small values of $\mathcal{R}_1$, because the Fourier spectrum of the wave function expands with such variations of $\Omega$ and $I_A$. For $I_A=0.0004$, $g_{max}\geq 6$ at $\Omega=0.1$ provides $\mathcal{R}_1 \leq 1.3\times 10^{-11}$, whereas $g_{max}\geq 16$ at $\Omega=1/1280$ provides $\mathcal{R}_1 \leq 2.0\times 10^{-6}$. For $\Omega=\Omega_A$, $g_{max}\geq 4$ at $I_A=3.9\times 10^{-7}$ provides $\mathcal{R}_1 \leq 7.1\times 10^{-15}$, whereas $g_{max}\geq 10$ at $I_A=0.0004$ provides $\mathcal{R}_1 \leq 8.6\times 10^{-11}$. Eq.~(\ref{a0aa0b}) is valid for the whole domain of study.

%%%%%%%%%%%%%%%%%%%%%%%%%%%%%figure2%%%%%%%%%%%%%%%
%%%%%%%%%%%%%%%%%%%%%%%%%%%%%%%%%%%%%%%%%%%%%%%%%%

%%%%%%%%%%%%%%%%%%%%%%%%%%%%%figure3%%%%%%%%%%%%%%%
%%%%%%%%%%%%%%%%%%%%%%%%%%%%%%%%%%%%%%%%%%%%%%%%%%

%%%%%%%%%%%%%%%%%%%%%%%%%%%%%figure4%%%%%%%%%%%%%%%
%%%%%%%%%%%%%%%%%%%%%%%%%%%%%%%%%%%%%%%%%%%%%%%%%%

%%%%%%%%%%%%%%%%%%%%%%%%%%%%%figure5%%%%%%%%%%%%%%%
%%%%%%%%%%%%%%%%%%%%%%%%%%%%%%%%%%%%%%%%%%%%%%%%%%

\subsection{Doublet structure of the ground state}
\begin{figure}
\includegraphics{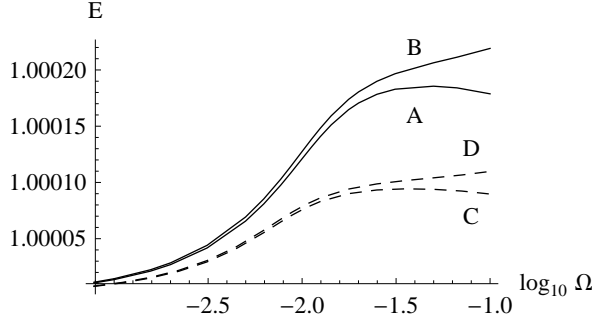}
\caption{\label{fig6}Normalized energy $E$ against $\log_{10}\Omega$ at (A) $I_A=0.0004$, $\xi=\xi_{0a}$, (B) $I_A=0.0004$, $\xi=\xi_{0b}$, (C) $I_A=0.0002$, $\xi=\xi_{0a}$, (D) $I_A=0.0002$, $\xi=\xi_{0b}$.}
\end{figure}

\begin{figure}
\includegraphics{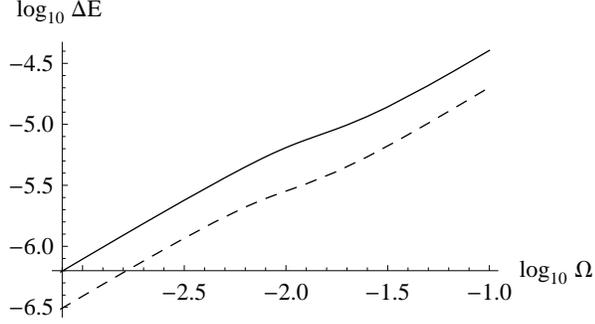}
\caption{\label{fig7}Logarithm of $\Delta E=E_b-E_a$ against $\log_{10}\Omega$ at $I_A=0.0004$ (solid curve) and $I_A=0.0002$ (dashed curve).}
\end{figure}

\begin{figure}
\includegraphics{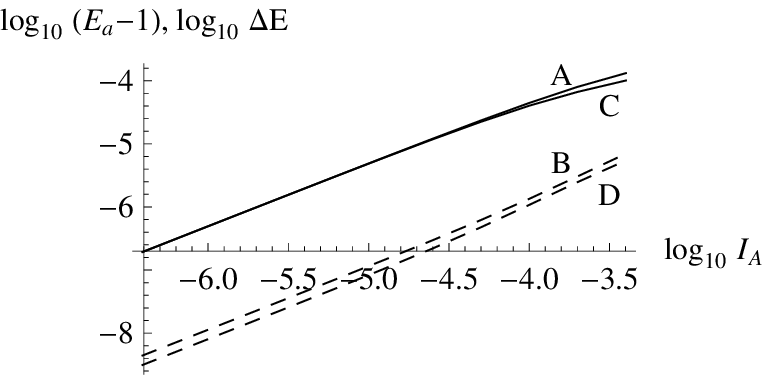}
\caption{\label{fig8}Plot of $\log_{10} (E_a - 1)$ against $\log_{10} I_A$ (solid curves) at (A) $\Omega=\Omega_A$, and (C) $\Omega=\Omega_B$. Plot of $\log_{10} \Delta E$ against $\log_{10} I_A$ (dashed curves) at (B) $\Omega=\Omega_A$, and (D) $\Omega=\Omega_B$.}
\end{figure}

Let us now compare the ground state wave functions specified by $\{\xi_{0a},a_{0a}\}$ and $\{\xi_{0b},a_{0b}\}$ in terms of the corresponding mean values of Hamiltonian
\begin{equation}\label{Hamilton}
    H=c\sum_{k=1}^3\alpha_k p_k + m_e c^2\alpha_4,
\end{equation}
operators of kinetic momentum
\begin{equation}\label{pk}
   p_k=-i\hbar\frac{\partial}{\partial x_k} - \frac{e}{c}A_k,
\end{equation}
probability current density $j_k=c\alpha_k$, and spin $S_k=\frac{\hbar}{2}\Sigma_k$, $k=1, 2, 3$. Both of these functions provide mean values: $\langle j_k \rangle =0$, $\langle p_k \rangle =0$, $k=1, 2, 3$, and $\langle S_2 \rangle=\langle S_3 \rangle=0$. The mean values ${\langle S_1\rangle}_a=\frac{\hbar}{2}{\langle\Sigma_1\rangle}_a$  and ${\langle S_1 \rangle}_b=\frac{\hbar}{2}{\langle\Sigma_1\rangle}_b$ for the doublet lines $a$ and $b$, respectively, are equal in magnitude but opposite in sign. They depend on $I_A$ and can be approximated as follows
\begin{equation}\label{S1apr}
    {\langle \Sigma_1 \rangle}_a=-{\langle \Sigma_1 \rangle}_b\approx 1 - I_A + \frac{3}{2}I_A^2.
\end{equation}
The normalized energy levels $E_a$ and $E_b$ of the doublet are different and depend on both $\Omega$ and $I_A$ as shown in Fig.~\ref{fig6}, Fig.~\ref{fig7}, and Fig.~\ref{fig8}, where $E= {\langle H \rangle}/(m_e c^2)$.

%%%%%%%%%%%%%%%%%%%%%%%%%%%%%figure6%%%%%%%%%%%%%%%
%%%%%%%%%%%%%%%%%%%%%%%%%%%%%%%%%%%%%%%%%%%%%%%%%%

%%%%%%%%%%%%%%%%%%%%%%%%%%%%%figure7%%%%%%%%%%%%%%%
%%%%%%%%%%%%%%%%%%%%%%%%%%%%%%%%%%%%%%%%%%%%%%%%%%

%%%%%%%%%%%%%%%%%%%%%%%%%%%%%figure8%%%%%%%%%%%%%%%
%%%%%%%%%%%%%%%%%%%%%%%%%%%%%%%%%%%%%%%%%%%%%%%%%%

\subsection{Spin precession}
The whole family of the ground state wave functions is defined the evolution operator (see Eqs.~(\ref{Ex}) and (\ref{phinx}))
\begin{eqnarray}\label{Ex1x4}
% \nonumber to remove numbering (before each equation)
  &&E_v(\bm{x})\equiv E_v(X_1,X_4) =\\
  e^{i \varphi_b}&&\left( \sum_{n\in S_{da}}S_a(n)P_a e^{i (\varphi'_n+\varphi_{ab})} + \sum_{n\in S_{db}}S_b(n)P_b e^{i \varphi'_n}\right),\nonumber
\end{eqnarray}
where $\varphi_b=-2 \pi (1+\xi_{0b})X_4/\Omega$, $\varphi_{ab}=2 \pi \Delta\xi X_4/\Omega$, and $\varphi'_n=2 \pi (n_1 X_1 - n_4 X_4)$, $S_{da}$ and $S_{db}$ are the solution domains of the doublet lines $a$ and $b$, respectively. In this case, the set $S_{da}$ contains only points $n=(n_1,0,0,n_4)\in\mathcal L$ with $n_4=-1, 0$, whereas the set $S_{db}$ contains only points $n$ with $n_4=0, 1$. The Fourier amplitudes $a=a(n)$ and $b=b(n)$ [see Eq.~(\ref{psiab})] have the following symmetry properties:
\begin{eqnarray}
% \nonumber to remove numbering (before each equation)
  a^*&=&(-1)^{n_4}a,\quad \Sigma_1 a=(-1)^{n_4}a,\quad n\in S_{da}, \\
  b^*&=&(-1)^{n_4}b,\quad \Sigma_1 b=-(-1)^{n_4}b,\quad n\in S_{db}.
\end{eqnarray}
Each member $\Psi=E_v(\bm{x})a_0$ of this family is specified by the amplitude $a_0$ which can be written without loss of generality as
\begin{equation}\label{a0delta}
    a_0=a_{0a}e^{i \delta}\cos\alpha + a_{0b}\sin\alpha,
\end{equation}
where $\alpha\in [0,\pi/2]$ and $\delta\in [0,2\pi]$.

\begin{figure}
\includegraphics{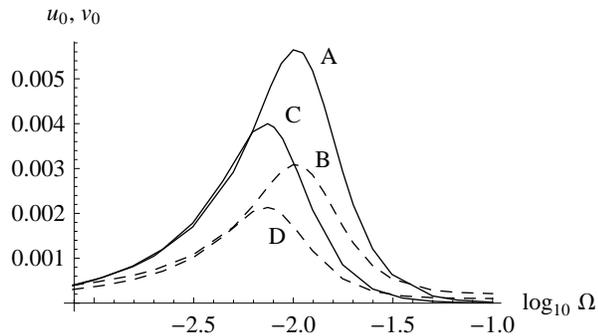}
\caption{\label{fig9}Plot of $u_0$ against $\log_{10}\Omega$ (solid curves) at (A) $I_A=0.0004$, and (C) $I_A=0.0002$. Plot $v_0$ against $\log_{10}\Omega$ (dashed curves) at (B) $I_A=0.0004$, and (D) $I_A=0.0002$.}
\end{figure}

\begin{figure}
\includegraphics{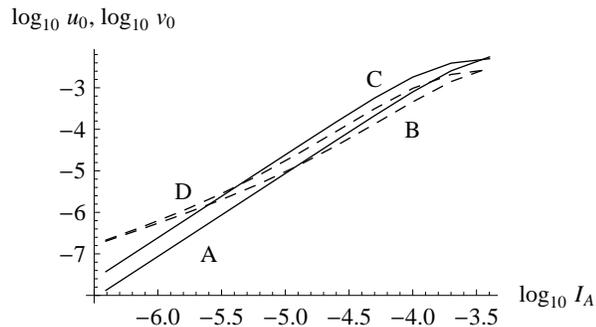}
\caption{\label{fig10}Plot of $\log_{10} u_0$ against $\log_{10} I_A$ (solid curves) at (A) $\Omega=\Omega_A$, and (C) $\Omega=\Omega_B$. Plot $\log_{10} v_0$ against $\log_{10} I_A$ (dashed curves) at (B) $\Omega=\Omega_A$, and (D) $\Omega=\Omega_B$.}
\end{figure}

The matrix function $E_v(X_1,X_4)$ is periodic in $X_1$. It is not periodic in $X_4$, but $\Delta\xi/\Omega\ll 1$, so that variations of $\varphi_{ab}$ at any unit interval of the $X_4$ axis are negligibly small, for example, in calculation of norms and mean values using Eqs.~(\ref{meanA}), (\ref{UE}) and (\ref{IntX}). In this approximation, for the normalized energy $E$ and the mean value $\langle\bm S\rangle=\frac{\hbar}{2}\langle\bm\Sigma\rangle$ of the spin operator one can readily obtain the relations:
\begin{equation}\label{Ealpha}
    E= {\langle H \rangle}/(m_e c^2)=E_a+\frac{\Delta E \sin^2\alpha}{1-u_0 \cos^2\alpha},
\end{equation}
\begin{equation}\label{spin}
    \langle\bm \Sigma\rangle=\frac{\textbf{e}_1{\langle \Sigma_1 \rangle}_a(\cos 2\alpha-u_0 \cos^2\alpha)+\textbf{e}_{\rho}(v_0-1)\sin 2\alpha}{1-u_0 \cos^2\alpha},
\end{equation}
where $\textbf{e}_{\rho}=\textbf{e}_3\cos\varphi+\textbf{e}_2\sin\varphi, \varphi=\delta+ 2\pi\Delta\xi X_4/\Omega=\delta+2\pi\nu_{pr}t$, $\delta$ specifies the initial precession phase,  $\nu_{pr}=\Delta\xi m_e c^2/h$ is the precession frequency, and
\begin{equation}\label{u0ab}
    u_0=1-\frac{u_a}{u_b}, u_a=\sum_{n\in S_{da}}|a(n)|^2, u_b=\sum_{n\in S_{db}}|b(n)|^2,
\end{equation}
\begin{equation}\label{sigma1a}
    {\langle \Sigma_1 \rangle}_a=\frac{1}{u_a}\sum_{n\in S_{da}}(-1)^{n_4}|a(n)|^2,
\end{equation}
\begin{eqnarray}
% \nonumber to remove numbering (before each equation)
  v_0&-&1=\frac{1}{u_b}\sum_{n\in S_{da}\bigcap S_{db}}a^{\dag}(n)\Sigma_3 b(n) \nonumber\\
  &=&\frac{2}{u_b}\sum_{n\in S_{da}\bigcap S_{db}}[a^1(n)b^1(n)+a^3(n)b^3(n)].
\end{eqnarray}
The mean values $\langle j_k \rangle$ and $\langle p_k \rangle$ ($k=1, 2, 3$) of the probability current density operators $j_k$ and the kinetic momentum operators $p_k$ are equal to zero for any ground state wave function $\Psi$. The mean value ${\langle \Sigma_1 \rangle}_a$ depend on $I_A$ and can be approximated by Eq.~(\ref{S1apr}), parameters $u_0$ and $v_0$ depend on $\Omega$ and $I_A$ as shown in Fig.~\ref{fig9} and Fig~\ref{fig10}, respectively.

%%%%%%%%%%%%%%%%%%%%%%%%%%%%%figure9%%%%%%%%%%%%%%%
%%%%%%%%%%%%%%%%%%%%%%%%%%%%%%%%%%%%%%%%%%%%%%%%%%

%%%%%%%%%%%%%%%%%%%%%%%%%%%%%figure10%%%%%%%%%%%%%%%
%%%%%%%%%%%%%%%%%%%%%%%%%%%%%%%%%%%%%%%%%%%%%%%%%%

Since $u_0\ll 1$ and $v_0\ll 1$, the ground state wave functions specified by $0\leq\alpha\leq\frac{\pi}{2}$ describe various spin states of the Dirac electron, including the spin precession with the frequency $\nu_{pr}$ at $0<\alpha<\frac{\pi}{2}$. The corresponding normalized energy levels $E$ fill the band from $E_a$ to $E_b=E_a + \Delta E$, see Fig.~\ref{fig6}, Fig.~\ref{fig7}, and Fig.~\ref{fig8}. The frequency $\nu_{pr}$ is defined by $\Delta\xi=\xi_{0b}-\xi_{0a}$, see Fig.~\ref{fig3} and Fig.~\ref{fig5}, in particular, $\nu_{pr}=3.062347 \times 10^{14}$ Hz at $\Omega=\Omega_A, I_A=0.0004$, and $\nu_{pr}=1.082908 \times 10^{14}$ Hz at $\Omega=\Omega_B, I_A=0.0002$.

Replacing the amplitudes $\textbf{A}_1=\textbf{A}_4$~(\ref{A1A4}) by $\textbf{A}_1=\textbf{A}_4=A_m(\textbf{e}_2-i \textbf{e}_3)/\sqrt{2}$ inverts the signs of ${\langle \Sigma_1 \rangle}_a$ and ${\langle \Sigma_1 \rangle}_b$ and reverses the precession direction, i.e., $\textbf{e}_{\rho}$ in Eq.~(\ref{spin}) takes the form $\textbf{e}_{\rho}=\textbf{e}_3\cos\varphi-\textbf{e}_2\sin\varphi$. In the case of counterpropagating waves with the same circular polarization ($\textbf{A}_1=\textbf{A}^*_4=A_m(\textbf{e}_2\pm i \textbf{e}_3)/\sqrt{2}$) or the same linear polarization ($\textbf{A}_1=\textbf{A}_4=A_m\textbf{e}_2$),  the spin precession is absent, because $\Delta\xi\equiv 0$.

\section{Conclusion}
The fundamental solution of the Dirac equation for an electron in the electromagnetic field with four--dimensional periodicity is obtained. The projection operator $\mathcal{S'}$~(\ref{SkbL}) defines the exact fundamental solution of the finite subsystem~(\ref{PkbLC}) which expands with each new step of the recurrent process. The relations, presented above and in~\cite{ESTCp1,ESTCp2}, form the complete set which is sufficient for the fractal expansion of this subsystem to a finite model of ESTC of any desired size. A criterion for evaluating accuracy of the approximate solutions, obtained by the use of such model, is suggested. It plays a leading role in search for the best approximate solutions in the framework of the selected model. The presented techniques are illustrated by analyzing the ground state of the Dirac electron in the field of counterpropagating plane waves. It is shown that in the electromagnetic lattice, composed by the left and right circularly polarized waves, the ground state is described by the family of wave functions with zero mean values of the probability current density operators and kinetic momentum operators, but with different energy levels and various spin states, including the spin precession.

\appendix*
\section{}
\subsection{Dirac basis for the linear space of $4\times 4$ matrices}
Let us enumerate 16 Dirac matrices, forming a basis for the linear space of $4\times 4$ matrices, by taking into account both interrelations between $2\times 2$ blocks of each matrix and interrelations between elements of each nonzero $2\times 2$ block as follows
\begin{equation*}
    \Gamma_0=\left(
    \begin{array}{cccc}
     1 & 0 & 0 & 0 \\
     0 & 1 & 0 & 0 \\
     0 & 0 & 1 & 0 \\
     0 & 0 & 0 & 1 \\
     \end{array}
     \right)=U,
\end{equation*}
\begin{equation*}
    \Gamma_1=\left(
    \begin{array}{cccc}
    1 & 0 & 0 & 0 \\
    0 & -1 & 0 & 0 \\
    0 & 0 & 1 & 0 \\
    0 & 0 & 0 & -1 \\
    \end{array}
    \right)=\Sigma_3,
\end{equation*}
\begin{equation*}
    \Gamma_2=\left(
    \begin{array}{cccc}
    0 & 1 & 0 & 0 \\
    1 & 0 & 0 & 0 \\
    0 & 0 & 0 & 1 \\
    0 & 0 & 1 & 0 \\
    \end{array}
    \right)=\Sigma_1,
\end{equation*}
\begin{equation*}
    \Gamma_3=\left(
    \begin{array}{cccc}
    0 & -i & 0 & 0 \\
    i & 0 & 0 & 0 \\
    0 & 0 & 0 & -i \\
    0 & 0 & i & 0 \\
     \end{array}
     \right)=\Sigma_2,
\end{equation*}
\begin{equation*}
    \Gamma_4=\left(
    \begin{array}{cccc}
    1 & 0 & 0 & 0 \\
    0 & 1 & 0 & 0 \\
    0 & 0 & -1 & 0 \\
    0 & 0 & 0 & -1 \\
    \end{array}
    \right)=\gamma_4=\alpha_4,
\end{equation*}
\begin{equation*}
    \Gamma_5=\left(
    \begin{array}{cccc}
    1 & 0 & 0 & 0 \\
    0 & -1 & 0 & 0 \\
    0 & 0 & -1 & 0 \\
    0 & 0 & 0 & 1 \\
    \end{array}
    \right)=\tau_3,
\end{equation*}
\begin{equation*}
    \Gamma_6=\left(
    \begin{array}{cccc}
    0 & 1 & 0 & 0 \\
    1 & 0 & 0 & 0 \\
    0 & 0 & 0 & -1 \\
    0 & 0 & -1 & 0 \\
    \end{array}
    \right)=\tau_1,
\end{equation*}
\begin{equation*}
    \Gamma_7=\left(
    \begin{array}{cccc}
    0 & -i & 0 & 0 \\
    i & 0 & 0 & 0 \\
    0 & 0 & 0 & i \\
    0 & 0 & -i & 0 \\
    \end{array}
    \right)=\tau_2,
\end{equation*}
\begin{equation*}
    \Gamma_8=\left(
    \begin{array}{cccc}
    0 & 0 & -1 & 0 \\
    0 & 0 & 0 & -1 \\
    -1 & 0 & 0 & 0 \\
    0 & -1 & 0 & 0 \\
    \end{array}
    \right)=\gamma_5,
\end{equation*}
\begin{equation*}
    \Gamma_9=\left(
    \begin{array}{cccc}
    0 & 0 & 1 & 0 \\
    0 & 0 & 0 & -1 \\
    1 & 0 & 0 & 0 \\
    0 & -1 & 0 & 0 \\
    \end{array}
    \right)=\alpha_3
\end{equation*}
\begin{equation*}
    \Gamma_{10}=\left(
    \begin{array}{cccc}
    0 & 0 & 0 & 1 \\
    0 & 0 & 1 & 0 \\
    0 & 1 & 0 & 0 \\
    1 & 0 & 0 & 0 \\
    \end{array}
    \right)=\alpha_1,
\end{equation*}
\begin{equation*}
    \Gamma_{11}=\left(
    \begin{array}{cccc}
    0 & 0 & 0 & -i \\
    0 & 0 & i & 0 \\
    0 & -i & 0 & 0 \\
    i & 0 & 0 & 0 \\
    \end{array}
    \right)=\alpha_2,
\end{equation*}
\begin{equation*}
    \Gamma_{12}=\left(
    \begin{array}{cccc}
    0 & 0 & i & 0 \\
    0 & 0 & 0 & i \\
    -i & 0 & 0 & 0 \\
    0 & -i & 0 & 0 \\
    \end{array}
    \right)=\tau_4,
\end{equation*}
\begin{equation*}
    \Gamma_{13}=\left(
    \begin{array}{cccc}
    0 & 0 & -i & 0 \\
    0 & 0 & 0 & i \\
    i & 0 & 0 & 0 \\
    0 & -i & 0 & 0 \\
    \end{array}
    \right)=\gamma_3,
\end{equation*}
\begin{equation*}
    \Gamma_{14}=\left(
    \begin{array}{cccc}
    0 & 0 & 0 & -i \\
    0 & 0 & -i & 0 \\
    0 & i & 0 & 0 \\
    i & 0 & 0 & 0 \\
    \end{array}
    \right)=\gamma_1,
\end{equation*}
\begin{equation*}
    \Gamma_{15}=\left(
    \begin{array}{cccc}
    0 & 0 & 0 & -1 \\
    0 & 0 & 1 & 0 \\
    0 & 1 & 0 & 0 \\
    -1 & 0 & 0 & 0 \\
    \end{array}
    \right)=\gamma_2.
\end{equation*}
Commonly used notation to the right of each matrix is given for convenience. At the presented numeration order, the structural information on each matrix $\Gamma_{\nu}$ is enclosed in its number $\nu$, i.e., one can reconstruct $\Gamma_{\nu}$ from $\nu$, and the multiplication rule for $\Gamma_{\lambda}\Gamma_{\mu}$ can be written as a function of $\lambda$ and $\mu$~\cite{ESTCp1}.

Any $4\times 4$ matrix $A$ can be written
\begin{equation*}
    A=\sum_{\nu=0}^{15}A_{\nu}\Gamma_{\nu},
\end{equation*}
where $A_{\nu}=\frac14 tr(A\Gamma_{\nu})$, and $tr\, A=4 A_0$. To single out the specific basis used in this expansion, the set of coefficients  $\{A_{\nu}\}$ is called in this article the Dirac set of matrix $A$, briefly, D-set of $A$, and it is denoted $D_s(A)$. This approach is of particular assistance in solving the system of Eqs.~(\ref{meq}). It is best suited to the structure of its matrix coefficients, accelerates numerical calculations and reduces data files. It should be emphasized that all major matrix operations (summation, multiplication, inversion, etc.) can be performed directly with D-sets, i.e., without matrix form retrieval~\cite{ESTCp1}.

\subsection{Projection operator of a system of homogeneous linear equations}
Let $\mathcal{V}$ and $\mathcal{V}^\ast$ be a linear space (finite or infinite dimensional) and its dual. At given $\omega \in \mathcal{V}^\ast$, the linear homogeneous equation in $\bm{x} \in \mathcal{V}$
\begin{equation}\label{omx}
    \langle\omega,\bm{x}\rangle = 0
\end{equation}
can be transformed to the equivalent equation
\begin{equation}\label{alx}
    \alpha\bm{x} = 0,
\end{equation}
where
\begin{equation}\label{aldyad}
    \alpha = \frac{\omega^\dag\otimes\omega}{\langle\omega,\omega^\dag\rangle}
\end{equation}
is the Hermitian projection operator (dyad) with the trace $tr\,\alpha=1$, and $\omega^\dag\in\mathcal{V}$. Let $U$ be the unit operator, i.e., $U\bm{x}=\bm{x}$ for any $\bm{x} \in \mathcal{V}$ and $\omega U = \omega$ for any $\omega \in \mathcal{V}^\ast$. The Hermitian projection operator $S=U-\alpha$ is the fundamental solution of (\ref{alx}), i.e., for any given $\bm{x}_0 \in \mathcal{V}$, $\bm{x}=S\bm{x}_0$ is a partial solution of (\ref{omx}) and (\ref{alx}).

Let now $\alpha$ and $\beta$ be Hermitian projection operators ($\alpha^\dag=\alpha^2=\alpha, \beta^\dag=\beta^2=\beta$) in $\mathcal{V}$. Providing the series
\begin{equation}\label{Aab}
    A=\alpha+\beta+\sum_{k=1}^{+\infty} \left[(\alpha\beta)^k\alpha-(\alpha\beta)^k+(\beta\alpha)^k\beta-(\beta\alpha)^k\right]
\end{equation}
is convergent, it defines the Hermitian projection operator with the following properties
\begin{eqnarray}
% \nonumber to remove numbering (before each equation)
  A^\dag=A^2=A,\quad \alpha A = A \alpha=\alpha,\nonumber\\
  \beta A = A \beta=\beta,\quad tr\,A=tr\,\alpha+ tr\,\beta. \label{Aprop}
\end{eqnarray}
Hence, the system of equations in $\bm{x}\in\mathcal{V}$
\begin{equation}\label{axbx}
    \alpha\bm{x}=0, \quad \beta\bm{x}=0
\end{equation}
reduces to one equation $A\bm{x}=0$ and has the fundamental solution $S=U-A$. The operator $A$ will be designated the projection operator of the system (\ref{axbx}). The trace $tr\,\alpha$ of the projection operator $\alpha$ specifies the dimension of the image $\alpha(\mathcal{V})$ of $\mathcal{V}$ under the mapping $\alpha$. It is significant that the relations (\ref{Aab}) and (\ref{Aprop}) are valid for any values of integers $tr\,\alpha$ and $tr\,\beta$. This enables us to extend this approach to systems with any (finite or infinite) number of homogeneous linear equations. To this end, we transform~(\ref{Aab}) to the following expression~\cite{bian04}
\begin{equation}
    A=(\alpha-\alpha\beta\alpha)^{-}(U-\beta)+(\beta-\beta\alpha\beta)^{-}(U-\alpha),\label{Apseu}
\end{equation}
where $(\alpha-\alpha\beta\alpha)^{-}$ is the pseudoinverse operator with the following properties
\begin{eqnarray}
% \nonumber to remove numbering (before each equation)
   &&(\alpha-\alpha\beta\alpha)^{-}(\alpha-\alpha\beta\alpha)=(\alpha-\alpha\beta\alpha)(\alpha-\alpha\beta\alpha)^{-}=\alpha,\nonumber\\
   &&\alpha(\alpha-\alpha\beta\alpha)^{-}=(\alpha-\alpha\beta\alpha)^{-}\alpha=(\alpha-\alpha\beta\alpha)^{-},\nonumber\\
   &&\sum_{k=1}^{+\infty}(\alpha\beta)^k=(\alpha-\alpha\beta\alpha)^{-}\beta.\label{pseu3}
\end{eqnarray}
The similar relations for $(\beta-\beta\alpha\beta)^{-}$ can be obtained from (\ref{pseu3}) by the replacement $\alpha\leftrightarrow\beta$. Numerical implementation of the pseudoinversion reduces to the inversion of $(tr\,\alpha)\times (tr\,\alpha)$ matrix for $(\alpha-\alpha\beta\alpha)^{-}$ and $(tr\,\beta)\times (tr\,\beta)$ matrix for $(\beta-\beta\alpha\beta)^{-}$.

In~\cite{bian04}, we have proposed a technique based on the use of (\ref{Apseu}) to find the fundamental solution of  the system~(\ref{PnC}). Here, we present the advanced version of this technique based on a fractal expansion of the system of equations taking into account and on the use of $A$ (\ref{Aab}) expressed as
\begin{equation}\label{Aad}
    A=\alpha+\delta, \quad \delta=(\beta-\alpha)\gamma(\beta-\alpha),
\end{equation}
where
\begin{equation}\label{gam}
    \gamma=\beta+\sum_{k=1}^{+\infty}(\beta\alpha\beta)^k=(\beta-\beta\alpha\beta)^{-},
\end{equation}
$\alpha,\beta,\delta$, and $A$ are projection operators, $\alpha,\beta,\gamma,\delta$, and $A$ are Hermitian operators interrelated as
\begin{eqnarray}
% \nonumber to remove numbering (before each equation)
  &&\beta\gamma=\gamma\beta=\gamma,\quad \beta\alpha\gamma=\gamma\alpha\beta=\gamma-\beta,\nonumber\\
  &&\alpha\delta=\delta\alpha=0,\quad \beta\delta=\beta-\beta\alpha,\quad \delta\beta=\beta-\alpha\beta,\nonumber\\
  &&\alpha A=A\alpha=\alpha, \quad \beta A=A\beta=\beta, \quad \delta A=A\delta=\delta. \nonumber
\end{eqnarray}
In the frame of this approach, calculation of all pseudoinverse operators in use reduces to the inversion of $4\times 4$ matrices.

\bibliography{BorzdovGN_2016}
\end{document}